# HATS-34b and HATS-46b: Re-characterisation Using TESS and Gaia

Emma M. Louden,[1]* Joel D. Hartman,[1]

[1]*Department of Astrophysical Sciences, Princeton University, NJ 08544, USA*

**ABSTRACT**

We present a revised characterisation of the previously discovered transiting planet systems HATS-34 and HATS-46. We make use of the newly available space-based light curves from the NASA *TESS* mission and high-precision parallax and absolute photometry measurements from the ESA *Gaia* mission to determine the mass and radius of the planets and host stars with dramatically increased precision and accuracy compared to published values, with the uncertainties in some parameters reduced by as much as a factor of seven. Using an isochrone-based fit, for HATS-34 we measure a revised host star mass and radius of $0.952^{+0.040}_{-0.020}\,M_\odot$ and of $0.9381 \pm 0.0080\,R_\odot$, respectively, and a revised mass and radius for the transiting planet of $0.951 \pm 0.050\,M_J$, and $1.282 \pm 0.064\,R_J$, respectively. Similarly, for HATS-46 we measure a revised mass and radius for the host star of $0.869 \pm 0.023\,M_\odot$, and $0.894 \pm 0.010\,R_\odot$, respectively, and a revised mass and radius for the planet of $0.158 \pm 0.042\,M_J$, and $0.951 \pm 0.029\,R_J$, respectively. The uncertainties that we determine on the stellar and planetary masses and radii are also substantially lower than re-determinations that incorporate the *Gaia* results without performing a full re-analysis of the light curves and other observational data. We argue that, in light of *Gaia* and *TESS*, a full re-analysis of previously discovered transiting planets is warranted.

## 1 INTRODUCTION

Our empirical knowledge about the physical properties of exoplanets comes mainly from observations of transiting exoplanets. Two particularly important parameters that can be measured for these objects are the planetary masses and radii. These are measured relative to the host star masses and radii, and thus the precision and accuracy of the planetary masses and radii are often limited by the precision with which the stellar parameters can be determined. Until very recently the great majority of transiting exoplanet host stars did not have geometric parallaxes available. In such cases the stellar masses and radii were determined by comparing the transit-inferred stellar density, and the spectroscopically determined temperature and metallicity of the star to theoretical stellar evolution models. This process typically yields stellar parameters that are accurate to ∼ 10 per cent. However, this has changed with the release of high-precision absolute parallax measurements for vast numbers of stars, including almost all known transiting planet host stars, by the *Gaia* mission (Gaia Collaboration et al. 2018). The parallax, together with absolute photometry, and the spectroscopically determined temperature, gives a direct measurement of the stellar radius, often to better than 1 per cent precision. This in turn, allows a much more precise measurement of the stellar mass as well.

Additionally, the NASA *TESS* mission (Ricker et al. 2015) is carrying out high-precision continuous photometric time-series observations for most of the sky. The mission will deliver the first space-based light curves for a large fraction of the known transiting planets with measured masses and radii. The new availability of the *Gaia* parallaxes and absolute photometry, and the newly available *TESS* light curves, provides an opportunity to significantly improve the precision of the masses and radii of known transiting exoplanets.

This paper will present the refined values for the planets HATS-34b (de Val-Borro et al. 2016) and HATS-46b (Brahm et al. 2018). These planets were discovered by the HATSouth Exoplanet Survey, which consists of six instruments at three sites around the globe that have been in operation since 2009 (Bakos et al. 2013). Since discovery, HATS-34b and HATS-46b have undergone follow-up observations by both *TESS* and *Gaia*.

In this paper we are focusing on HATS-34 and HATS-46 as the first two HATSouth planets observed by *TESS* (in Sector 1) which were discovered and analysed prior to the availability of *Gaia* DR2. Bulk re-determinations of planetary system parameters using *Gaia* DR2 have been published by Berger et al. (2018) for Kepler systems and Johns et al. (2018), for non-Kepler systems, but these works did not re-analyse the light curves and radial velocity data, and thus do not capture the full potential benefit of knowing the parallax that comes from enabling a tighter constraint on the stellar density and hence impact parameter and planetary radius. As we demonstrate in this paper, it is necessary to re-fit all of the data for each system, including the parallax constraints, to take full advantage of the available information, and achieve higher precision. Significant improvements in stellar and planetary parameters when incorporating parallax measurements have also been reported for the system KELT-11 (Beatty et al. 2017) and for several M dwarfs with planets observed by Kepler (Stevens et al. 2018). These works differ from the approach presented here in that they focused on measuring empirical stellar masses and radii, without relying on stellar evolution models. Here we present both empirical stellar parameters for HATS-34 and HATS-46, and parameters that rely on imposing constraints from stellar evolution models. As we show, the stellar evolution models still provide more precise parameter estimates than can be obtained from the purely empirical techniques. Finally, we note that Stevens et al. (2018) presented analytical formulae for estimating the expected uncertainties when analysing a transiting planet system with a precise parallax measurement. We compare our results to these analytic expectations in Section 5.4.

Section 2 will summarise the observations of these systems that we include in our analysis. Section 3 of this paper will describe our methods. Section 4 will present the results of the model fitting. Section 5 discusses the results.

## 2   OBSERVATIONS

### 2.1   Data from discovery papers

Both HATS-34 and HATS-46 previously underwent photometric and spectroscopic observations. The details of the observations can be found in de Val-Borro et al. (2016) and Brahm et al. (2018), respectively. We gathered the observational data presented in these papers, and incorporated them into our analyses of these two systems. For HATS-34 these observations include the HATSouth discovery light curve, follow-up light curves from the PEST 0.3 m telescope and from the DK 1.54 m, radial velocities (RVs) from FEROS on the MPG 2.2 m (Kaufer & Pasquini 1998), and the stellar effective temperature and metallicity determined from the FEROS spectra. For HATS-46 these observations include the HATSouth discovery light curve, follow-up light curves from the Swope and LCOGT 1 m telescopes (Brown et al. 2013), RVs from FEROS and PFS on the Magellan 6.5 m (Crane et al. 2010), and the stellar effective temperature and metallicity determined from the FEROS spectra.

### 2.2   TESS and Gaia

*TESS* (Ricker et al. 2015) and *Gaia* (Gaia Collaboration et al. 2018) are two space-based missions carrying out (almost) all-sky surveys and providing high-precision light curves and parallax measurements, respectively, for almost all the stars with known transiting planets. These new data will enable much higher precision measurements of the physical parameters of these stars and their planets.

The *TESS* observations of HATS-34 and HATS-46 (both of which were known prior to the launch of the mission) were carried out in short-cadence mode. We made use of the PDC light curves of these stars produced by the *TESS* Science Processing Operations

Centre (SPOC) at NASA Ames, which we downloaded from the Barbara A. Mikulski Archive for Space Telescopes (MAST)[1].

For HATS-34 we used this light curve directly in the fit, but for HATS-46 additional filtering of stellar variability was necessary. To do this filtering for HATS-46 we fit and subtracted a 20-harmonic Fourier series at a period of 27.88 days (equal to the time-span of the light curve). This is shown in Figure 3. In fitting the *TESS* light curves we allowed for a dilution term to be fit in a similar fashion to our treatment of the HATSouth light curves (as described in the discovery papers). The *TESS* light curves show no detectable occultations of the planets by the host stars, and also show no evidence for out-of-transit phase modulation at the orbital period of the system. The constraints on these effects are not much better than is possible from the HATSouth observations, so we do not attempt to quantify them here.

We searched for additional transits in the TESS light curve residuals using the Box Least Squares (BLS; Kovács et al. 2002) algorithm, and estimated the photometric rotation period of the stars from the TESS data using the Generalised Lomb-Scargle (GLS; Zechmeister & Kürster 2009) algorithm together with a bootstrap procedure to calibrate the false alarm estimates. We used the VAR-TOOLS (Hartman & Bakos 2016) implementation for both of these. For HATS-34 we found marginal evidence for a rotation period of 6.484 days with a formal false alarm probability of 0.3 per cent and no significant sign of additional transits. Though not discussed in the discovery paper, the HATSouth light curve for HATS-34 shows no detectable periodic signal that can be associated with the rotation period of the star. For HATS-46 we found a clear periodic modulation at a period of 15.487 days and no significant evidence for additional transits. If the variability seen in the *TESS* light curve were real, we would expect to have seen the signal in the HAT-South light curve as well, unless the amplitude of the variation had increased significantly between the time of the HATSouth observations and the *TESS* observations. Such an increase cannot be ruled out, as active stars are known to show secular evolution in their variability over time. As for HATS-34, the HATSouth light curve for HATS-46 shows no evidence for periodic variability that could be associated with the rotation of the star.

In both cases the light curve artefacts from momentum dumps and bad DQ flags were removed prior to analysis.

We also made use of the *Gaia* parallax, and *G*-, *BP*- and *RP*-band photometry for each star, as listed in the *Gaia* DR2 catalog. Additionally we included broad-band Infrared (IR) photometry from the WISE mission (Cutri et al. 2012), which was not included in the prior analyses of these systems.

## 3   METHODS

For each system joint modelling was carried out using the light curves (original, follow-up, and *TESS*), radial velocities, stellar atmospheric parameters, broad-band photometry, and *Gaia* parallax observations following the method described in Hartman et al. (2019). We fit the observations using a Differential Evolution Markov Chain Monte Carlo (DEMCMC) procedure, producing over 300,000 links for both systems. We visually inspected the chains to ensure convergence and determine the burn-in period to exclude in calculating the posterior parameter estimates and uncertainties.

---

[1] https://archive.stsci.edu/tess/index.html

Table 1 lists the parameters that we varied in the fit and the associated priors. Corner plots showing the correlations between varied parameters are provided in Appendix A.

We determined the stellar masses and radii following two different methods. The first of these is a mostly empirical method and the second relies on comparison to theoretical stellar evolution models.

### 3.1 Transit Fitting

In this work a joint analysis was performed of the RV observations, light curves, Gaia DR2 parallax, catalog broad-band photometry from Gaia, UCAC4, 2MASS and WISE, and the values for $T_{\rm eff\star}$ and [Fe/H] from the ZASPE analysis of the spectra.

Light curves were modelled using the semi-analytic relations from Mandel & Agol (2002), assuming quadratic limb darkening. The limb darkening coefficients were allowed to vary in the fit, with Gaussian priors imposed on their values using the theoretical tabulations from Claret et al. (2012, 2013); Claret (2018). The uncertainties for each light curve were scaled to achieve a $\chi^2$ per degree of freedom of one about the best-fit model. For HATSouth and *TESS* we included dilution factors in the fit which allow the transit depth to be reduced due to blending with neighbors, or errors in the trend-filtering process (which is performed prior to fitting the transit model for these light curves). As noted in the discovery papers, the HAT-South data were detrended using the Trend Filtering Algorithm per Kovács et al. (2005). A well-known side effect of this filter is that it tends to distort and dilute transit signals and thus can be accounted for in the fit by the previously noted dilution factors. This is a common procedure that is applied to all HATS systems. In practice we found that for both HATS-34 and HATS-46 the *TESS* light curves were consistent with no effective dilution. Since the *TESS* SPOC pipeline subtracts the flux of known neighbours from the light curves, the lack of dilution in the modelling is not surprising. For the ground-based follow-up light curves we allowed for a linear and quadratic trend in time in the out-of-transit magnitude, and we also allowed for linear trends in three parameters used in fitting a rotated ellipsoidal Gaussian to the the point-spread-function of each image. These trend terms were fit simultaneously to the transit model.

The RV observations were modelled using the analytic relations for Keplerian orbits. Jitter terms were added in quadrature to the RV uncertainties for each instrument, and were allowed to vary in the fit. We performed each joint analysis twice, first assuming a fixed circular orbit, and then allowing the eccentricity to vary in the fit (specifically we used $\sqrt{e}\cos\omega$ and $\sqrt{e}\sin\omega$ as the free parameters). The motivation for considering a fixed circular orbit is that tidal forces are expected to quickly circularise the orbits of these short period planets, and this expectation has been borne out by secondary eclipse and high-precision RV observations for several hot Jupiters with periods and masses comparable to HATS-34b and HATS-46b (e.g., Dawson & Johnson 2018). Allowing the eccentricity to vary in the analysis when the orbit is circular leads not only to larger errors in the parameters, it can also lead to systematic errors in the values estimated from the median of the DEMCMC posterior distributions (e.g., Hartman et al. 2011).

We modelled the broad-band catalog photometry, parallax, and spectroscopic parameters by introducing the distance modulus, the *V*-band extinction, $T_{\rm eff\star}$, and [Fe/H] as additional free parameters in the fit, and then either (a) allowing the stellar radius to also vary, and using theoretical bolometric correction tables (Section 3.2), or (b) using theoretical stellar evolution models to constrain the relations between various stellar physical parameters (Section 3.3). In both cases the light curve and RV observations also inform this modelling by constraining the stellar density $\rho_\star$ (we do not vary this parameter directly, but instead was found from the other orbital and transit parameters that were allowed to vary). For the extinction parameter $A_V$ we impose a Gaussian prior centred on the value estimated using the MWDUST Galactic extinction model (Bovy et al. 2016). We assume a Cardelli et al. (1989) extinction law, with $R_V = 3.1$ to determine the extinction in other band-passes given $A_V$.

After an initial analysis we found that the residuals for the broad-band photometry were much larger than the uncertainties, particularly for the *Gaia* DR2 observations which had formal uncertainties of a few mmag, including the estimated systematic error from Evans et al. (2018). We therefore choose to adopt a minimum uncertainty of 0.02 mag for all band-passes. This is roughly the level at which theoretical filter transmission profiles and stellar atmospheric models are thought to be reliable, at least for solar-type stars. Starspots, plage and other stellar activity phenomena likely also impact stellar photometry below this level of precision.

### 3.2 Empirical Determination of Stellar Mass and Radius

The mostly empirical method that we followed for measuring the stellar mass and radius is similar to that suggested by Stassun et al. (2017). For the remainder of the paper this method will be referred to as the empirical method, although we recognize that it relies upon some theoretical models and thus is not purely empirical.

The values for $R_p/R_\star$, the transit duration $T_{14}$, the impact parameter $b$, and $P$ from transit observations and $K$, $e$, and $\omega$ from radial velocity measurements are combined to get $M_p/M_\star^{2/3}$ and $\rho_s$. The stellar radius $R_\star$ is a free parameter in the fit, which is constrained by the parallax, stellar effective temperature, broad-band photometry, and bolometric corrections taken from the PARSEC isochrones (Marigo et al. 2017). The stellar mass is determined by combining $\rho_s$ and $R_s$. In practice we jointly fit the light curves, RV observations, broad-band photometry, spectroscopically determined stellar atmospheric parameters, and parallax, using $T_{\rm eff\star}$, [Fe/H], $R_\star$, the distance modulus, and $A_V$ as free parameters that describe the properties of the star.

The filter profiles assumed by the PARSEC models in producing the bolometric corrections are Maíz Apellániz & Weiler (2018) for Gaia DR2, Maíz Apellániz (2006) for UBV, Cohen et al. (2003) for the 2MASS filters, Fukugita et al. (1996) for the SDSS filters, and Wright et al. (2010) for the WISE filters.

### 3.3 Theoretical Stellar Model Determination of Stellar Mass and Radius

In this case the relevant measured stellar parameters (radius, temperature, metallicity, density, and absolute magnitudes in various band-passes) are compared to the PARSEC (Marigo et al. 2017) and MIST v1.2 (Dotter 2016; Choi et al. 2016; Paxton et al. 2011, 2013, 2015) theoretical stellar evolution models. We perform tri-linear interpolation within tabulated isochrones using the metallicity, effective temperature, and stellar density as the independent variables. The metallicity and effective temperature are varied directly in our joint fit, while the density is determined from the varied parameters used to model the transit light curves and RV observations. Constraining the stellar parameters to agree with theoretical models leads to tighter constraints on the stellar masses and radii compared to the empirical method. However, the uncertainties determined in this method do not account for possible systematic errors in the theoretical models, which may be larger than the uncertainties estimated from the fit. This issue is discussed further in Section 5.2. Figure 6 compares the *Gaia* DR2 broad-band photometry for HATS-34 and HATS-46, respectively, to the PARSEC theoretical isochrones projected onto the $G$ vs. $BP - RP$ diagram. These figures also compare the spectral-energy distributions (SEDs) for each star, observed in several broad-band filters, to theoretical SEDs sampled from the Markov Chains generated in fitting the observations.

**Table 1.** Parameters varied in joint analysis

| Parameter | Prior | Notes |
|---|---|---|
| $T_A$ | uniform | mid transit time of first observed transit |
| $T_B$ | uniform | mid transit time of last observed transit |
| $K$ | uniform, $K > 0$ | RV semi-amplitude |
| $\sqrt{e}\cos\omega$ | uniform, $0 \le e < 1$ | eccentricity parameter, either fixed to zero or varied |
| $\sqrt{e}\sin\omega$ | uniform, $0 \le e < 1$ | eccentricity parameter, either fixed to zero or varied |
| $R_p/R_\star$ | uniform | ratio of planetary to stellar radius |
| $b^2$ | uniform, $b^2 \ge 0$ | impact parameter squared |
| $\zeta/R_\star$ | uniform | reciprocal of the half duration of the transit |
| $\gamma_i$ | uniform | systemic velocity for RV instrument $i$ |
| $\sigma_{\text{jit,i}}$ | $-\log(\sigma_{\text{jit,i}}), \sigma_{\text{jit,i}} > 0$ | jitter for RV instrument $i$ |
| $c_{1,i}$ | Gaussian with $\sigma = 0.2$ mean based on Claret et al. (2012, 2013); Claret (2018) | First quadratic limb-darkening coefficient for filter $i$ |
| $c_{2,i}$ | Gaussian with $\sigma = 0.2$ mean based on Claret et al. (2012, 2013); Claret (2018) | Second quadratic limb-darkening coefficient for filter $i$ |
| $m_{0,HS,i}$ | uniform | out-of-transit magnitude for HS or *TESS* light curve $i$ |
| $d_{HS,i}$ | uniform, $0 < d_{HS,i} \le 1$ | transit dilution factor for HS or *TESS* light curve $i$ |
| $m_{0,LC,i}$ | uniform | out-of-transit magnitude for follow-up light curve $i$ |
| $m_{1,LC,i}$ | uniform | linear trend to out-of-transit magnitude for follow-up light curve $i$ |
| $m_{2,LC,i}$ | uniform | quadratic trend to out-of-transit magnitude for follow-up light curve $i$ |
| $S_{0,LC,i}$ | uniform | EPD coefficient for PSF shape parameter $S$ for follow-up light curve $i$ |
| $D_{0,LC,i}$ | uniform | EPD coefficient for PSF shape parameter $S$ for follow-up light curve $i$ |
| $K_{0,LC,i}$ | uniform | EPD coefficient for PSF shape parameter $S$ for follow-up light curve $i$ |
| $d_{\text{mod}}$ | $2\ln(\frac{d_{\text{mod}}+5}{5}) - \frac{d_{\text{mod}}+5}{7650}$ Gaussian with $\sigma = 0.025$ mag | distance modulus |
| $A_V$ | mean based on MWDUST model $A_V \ge 0$ | extinction |
| $T_{\text{eff}}$ | uniform, $T_{\text{eff}} > 0$ | host star effective temperature |
| [Fe/H] | uniform | host star metallicity |
| $R_\star$ | $\log R_\star, R_\star > 0$ | host star radius, only used for method in Section 3.2, not for method in Section 3.3 |

The sources for the filter profiles assumed by the PARSEC models are listed in Section 3.2. For MIST we are using Vega magnitudes for all filters and Evans et al. (2018) for Gaia DR2, Bessell & Murphy (2012) for UBV, Cohen et al. (2003) for 2MASS, SDSS DR7 [2] for SDSS filters, and Wright et al. (2010) for the WISE filters.

### 3.4 Excluding Blend Scenarios

A blend analysis following Hartman et al. (2019) was carried out to exclude blend scenarios for both systems. While such an analysis was also performed in the discovery papers for each object, the new *TESS* and *Gaia* DR2 data motivates us to revisit the issue in order to confirm the previous result that neither object is a blended stellar eclipsing binary system.

Modelling was done for four possible scenarios:

(i) HP- modelling the system as a single star with a transiting planet

(ii) HPS- modelling the system as a star with a transiting planet and an unresolved binary star companion

(iii) HSS- modelling the system as a hierarchical triple star system

(iv) HSSBGEB- modelling the system as a blend between a foreground star and a background eclipsing binary star system

The blend modelling showed that the HP, HPS, and HSS-BGEB models all provide comparably good fits for both systems, with slight improvements in $\chi^2$ for the HPS and HSSBGEB models compared to the HP model which are consistent with the improvement expected from additional free parameters to the model. The HSSBGEB model was excluded as a model by simulating spectral line bisector span (BS) and RV measurements for 1000 links from the HSSBGEB Markov chain and verifying that the simulations tested yielded BS scatter or RV variations well in excess of what was measured. For the HPS model we place a 95% confidence upper limit on the mass of an unresolved stellar companion of $0.32\,M_\odot$ for HATS-34, and $0.43\,M_\odot$ for HATS-46.

## 4 RESULTS

### 4.1 HATS-34

Figures 1, 2, 3, 4, 5 and 6 show the best-fit model for HATS-34 compared to the HATSouth light curve, ground-based follow-up light curves, de-trended *TESS* light curves, *TESS* light curves, RV observations, and broad-band catalogue photometry, respectively. These figures show the models from our MIST theoretical isochrone-based fitting, assuming a single star with a transiting planet on a circular orbit. Table 2 lists the derived stellar and planetary parameters determined from our modelling and the remaining orbital parameters can be found in Table 6. We list separately the parameters determined from the empirical stellar parameter method (Section 3.2), and the theoretical isochrone-based methods (Section 3.3), using the PARSEC and MIST stellar evolution models.

HATS-34 is a grazing system. As a result, the stellar density is poorly constrained from the light curves alone, leading to large uncertainties on the stellar mass (and hence planetary mass as well), extending well beyond the physically allowed range, when using the empirical method for determining the stellar parameters. For this system we suggest adopting the parameters determined using the MIST theoretical isochrones. Our adopted parameters include a stellar mass and radius of $0.952^{+0.040}_{-0.020} M_\odot$ and $0.9381 \pm 0.0080 R_\odot$, respectively, and a planetary mass and radius of $0.951 \pm 0.050 M_J$ and $1.282 \pm 0.064 R_J$, respectively.

Between the PARSEC isochrone model and the empirical model, we find that $a/R_\star$ differs by about $4\sigma$[3], with the empirical model preferring a lower value for $\frac{a}{R_\star}$ and a higher value for the impact parameter $b$. This difference then leads to large differences for other parameters that depend on these parameters, such as the stellar mass and the planetary mass.

The large differences between the empirical and isochrone-based modelling for these parameters is due to the grazing transits, which, as noted, can lead to large systematic errors in the inferred stellar density, and hence in the stellar and planetary masses and radii. For grazing transits there is a strong degeneracy between $a/R_\star$, $b$, and $R_p/R_\star$. The errors on these parameters can be underestimated in the MCMC analysis for a grazing system, leading to apparent discrepancies between the empirical and isochrone-based results that exceed the listed uncertainties, as is the case here. For HATS-34 the stellar radius, that is largely constrained by *Gaia*, together with the stellar effective temperature gives a much better constraint on the stellar mass (and hence density) when using theoretical isochrones to constrain the stellar parameters.

As noted in the discovery paper by de Val-Borro et al. (2016), the FEROS observations show fairly substantial bisector span variations that may either be attributed to sky contamination for this somewhat faint target, or the presence of an unresolved stellar companion. Based on our blend modelling, any such unresolved companion must have a mass less than $0.32 M_\odot$.

Compared to the values from the modelling as described in Section 3.3 using the PARSEC isochrones, the MIST isochrones produced values that are different by less than 1 $\sigma$ for the planetary mass, $1\sigma$ for the planetary radii, 1 $\sigma$ for stellar mass and 1 $\sigma$ for the stellar radii when eccentricity was allowed to vary. When constrained to a circular orbit the MIST isochrones produced values that vary by less than 1 $\sigma$ for the planetary mass, 1 $\sigma$ for the planetary radii, 1 $\sigma$ for the stellar mass, and 3 $\sigma$ for the stellar radii. The $\sim 3\%$ differences in the inferred stellar radii between the MIST and PARSEC isochrone values, which appears to stem at least in part from differences in the bolometric corrections used by these models, provides a measure of the systematic error between different stellar models for a $\sim 0.8 M_\odot$ star. We conclude that in this case we are limited by the accuracy of stellar models in our ability to determine the radius of this star. Despite the fact that the stellar radius measurements differ by more than the internal errors determined for each model, the planetary masses and radii are consistent at the $\sim 1\sigma$. level, suggesting that the systematic errors in these parameters due to systematic uncertainties in the stellar evolution models are comparable to the uncertainties in these parameters that arise from the precision of the observations included in the analysis. Higher precision observations are unlikely to further improve the planetary mass and radius measurements without concomitant improvements to the stellar models.

### 4.2 HATS-46

Figures 1, 2, 3, 5, 6, and 7 show the best-fit model for HATS-46 compared to the HATSouth light curve, ground-based follow-up light curves, de-trended *TESS* light curves, RV observations, broad-band catalogue photometry, and *TESS* light curve, respectively. These figures show the models from our MIST theoretical isochrone-based fitting, assuming a single star with a transiting planet on a circular orbit. Table 3 lists the derived stellar and planetary parameters determined from our modelling and the remaining orbital parameters can be found in Table 7. We list separately the parameters determined from the empirical stellar parameter method (Section 3.2), and the theoretical isochrone-based methods (Section 3.3) using the PARSEC and MIST stellar evolution models.

As for HATS-34, we find that the empirical model and PARSEC isochrone models yield differences that exceed the formal uncertainties for some parameters. For example, $\frac{a}{R_\star}$ differs by about $2.3\sigma$ between the two different models for HATS-46 (when constrained to e = 0). In this case the transits are not grazing, and the empirical method yields a more reliable result than for HATS-34. Nonetheless, the *Gaia* stellar radius, coupled with stellar evolution models, provides a much tighter constraint on the stellar mass and density than can be determined directly from the light curves. We again suggest adopting the isochrone-based parameters for this system. Our adopted parameters, which come from the MIST isochrones, include a stellar mass and radius of $0.869 \pm 0.023 M_\odot$ and $0.894 \pm 0.010 R_\odot$, respectively, and a planetary mass and radius of $0.158 \pm 0.042 M_J$ and $0.951 \pm 0.029 R_J$, respectively.

Compared to the values from the modelling as described in Section 3.3 using the PARSEC isochrones, the MIST isochrones produced values that are different by 1 $\sigma$ for the planetary mass, 1 $\sigma$ for the planetary radii, 2 $\sigma$ for the stellar mass, and 3 $\sigma$ for the stellar radii when eccentricity was allowed to vary. When constrained to a circular orbit the MIST isochrones produced values that vary by 1 $\sigma$ for the planetary mass and radii, 2 $\sigma$ for the stellar mass and a little over $2\sigma$ for the stellar radii.

---

[3] Here and elsewhere in the paper, in comparing posterior parameter values from different models, the $\sigma$ used for comparison is the quadrature sum of the uncertainties from each of the two parameters being compared. For parameters with two-sided uncertainties, we use the upper uncertainty when comparing to another parameter that has a greater value, and the lower uncertainty when comparing to another parameter that has a lesser value.

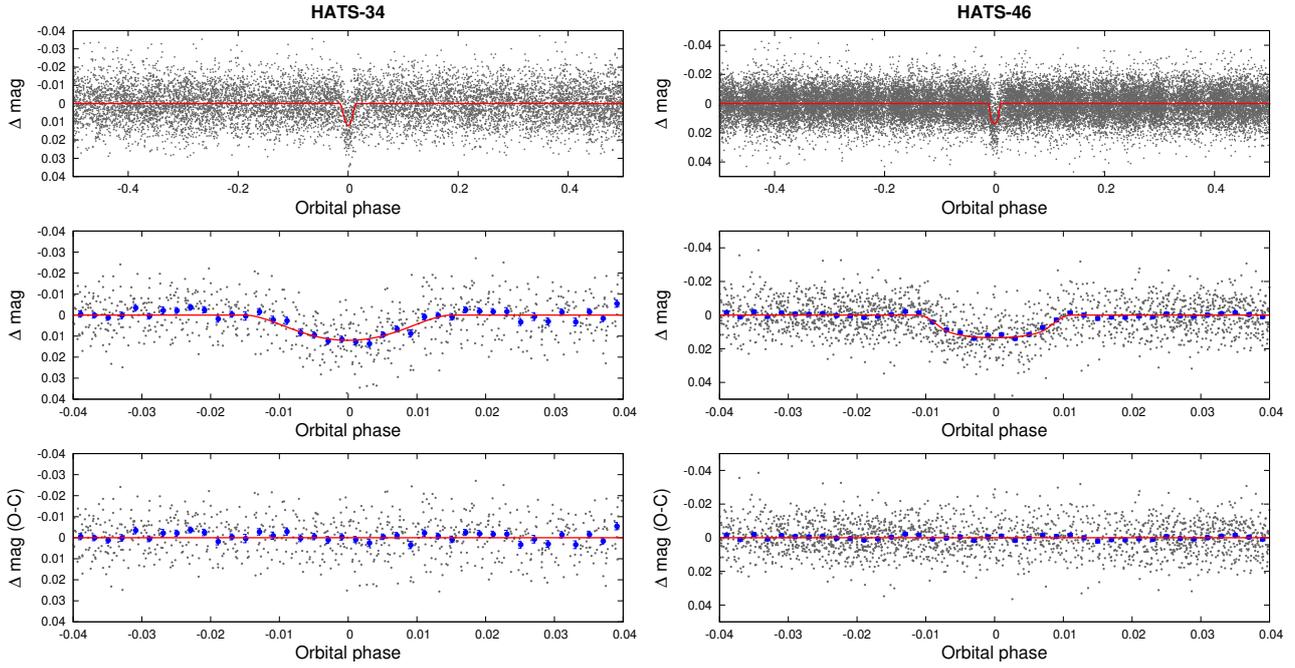

**Figure 1. HATSouth Light Curves of HATS-34 (*left*) and HATS-46 (*right*).** In each case the *top* panel shows the full light curve, the *middle* panel is zoomed in to show the transit in greater detail, and the *bottom* panel shows the residuals from the best-fit model. The solid line represents the model and the dark filled circles on the lower panels show the light curves binned in phase with a bin size of 0.002.

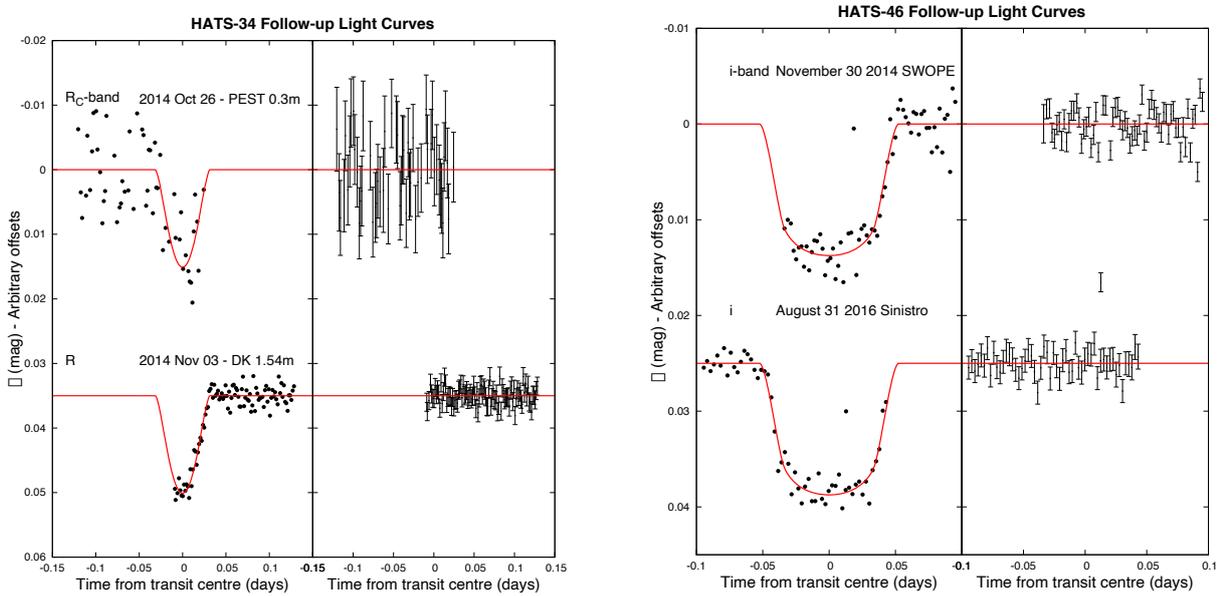

**Figure 2. Ground-based Follow-Up Light Curves of HATS-34 (*left*) and HATS-46 (*right*).** On the left side of each panel, the two follow-up light curves are shown with the models overlaid. On the right side of each panel are the residuals from the best-fit models in the same order as the original light curves are presented. The filters, dates, and instruments are as indicated on the figure.

## 5    DISCUSSION

We carried out a re-analysis of the transiting planet systems HATS-34 and HATS-46, including newly available data from *Gaia* and *TESS*. As we discuss below, we measure the stellar and planetary masses and radii to substantially higher precision than previous determinations.

### 5.1    Summary of Fitting Methods and Differences from Prior Work

As discussed in Section 3, and shown in Tables 2, 3, 6 and 7, we analysed each system using three distinct methods. For each analysis method we fitted the observations twice: once assuming a circular orbit for the planet, and once allowing for a non-zero eccentricity.

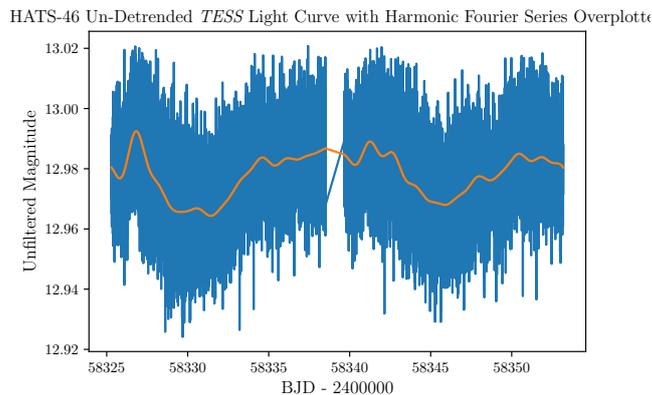

**Figure 3. Un-detrended Light Curve of HATS-46.** This plot shows the un-phased and un-detrended *TESS* light curves. For HATS-46, the 20-harmonic Fourier series is over plotted.

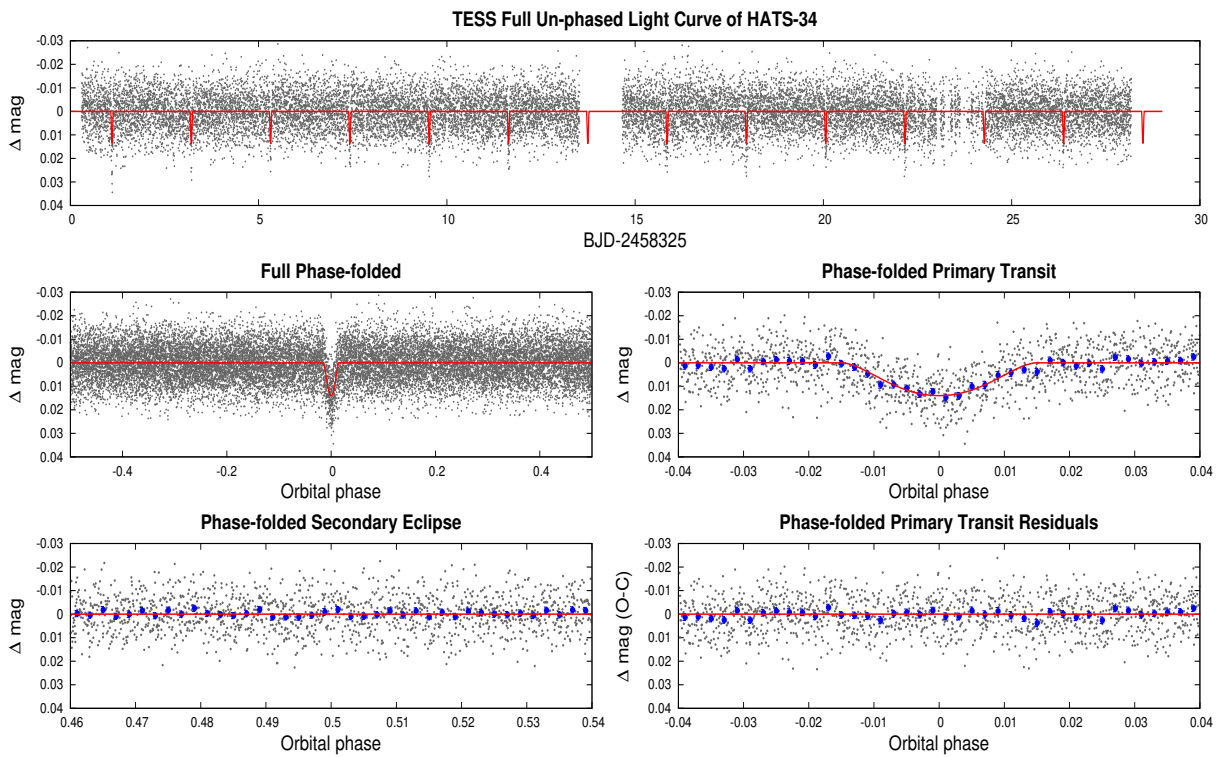

**Figure 4. TESS Light Curve of HATS-34.** The *top* plot shows the full light curve (with the model in red). The *middle* row of plots shows the full light curve after it has been phase folded (model in red) and a zoomed in plot of the primary transit with the model (red line) and the light curve binned in phase with a bin size of 0.002 (blue dots) overlaid. The *bottom* row shows the phase folded light curve zoomed in on the secondary eclipse with the model in red and the phase-folded residuals zoomed in on the primary transit.

This leads to a total of six different fits that were performed for each system. The three distinct fitting methods include: (1) an empirical method for constraining the stellar mass and radius from the transit-derived stellar density, and the effective temperature, bolometric flux and distance (the "PARSEC Empirical" method labelled in the tables); (2) an isochrone-based method that uses the PARSEC theoretical stellar evolution models to constrain the stellar properties from the observables (the "PARSEC Isochrone" method); and (3) a similar isochrone-based method that uses the MIST theoretical stellar evolution models (the "MIST Isochrone" method). We find that both systems have circular (or very nearly circular) or-bits, and that allowing the eccentricity to vary has only a modest impact on the resulting planetary masses and radii when imposing constraints from stellar evolution models. We also find that imposing the theoretical isochrone-based constraints generally improves the precision with which the stellar and planetary masses and radii are determined compared to the empirical method, especially when eccentricity is allowed to vary. We find that the planetary masses and radii derived from the MIST and PARSEC-based isochrones are comparable, but there are ∼ 1 − 3σ differences in the stellar masses and radii between these methods. This indicates that the accuracy of the stellar parameters for these two systems is now

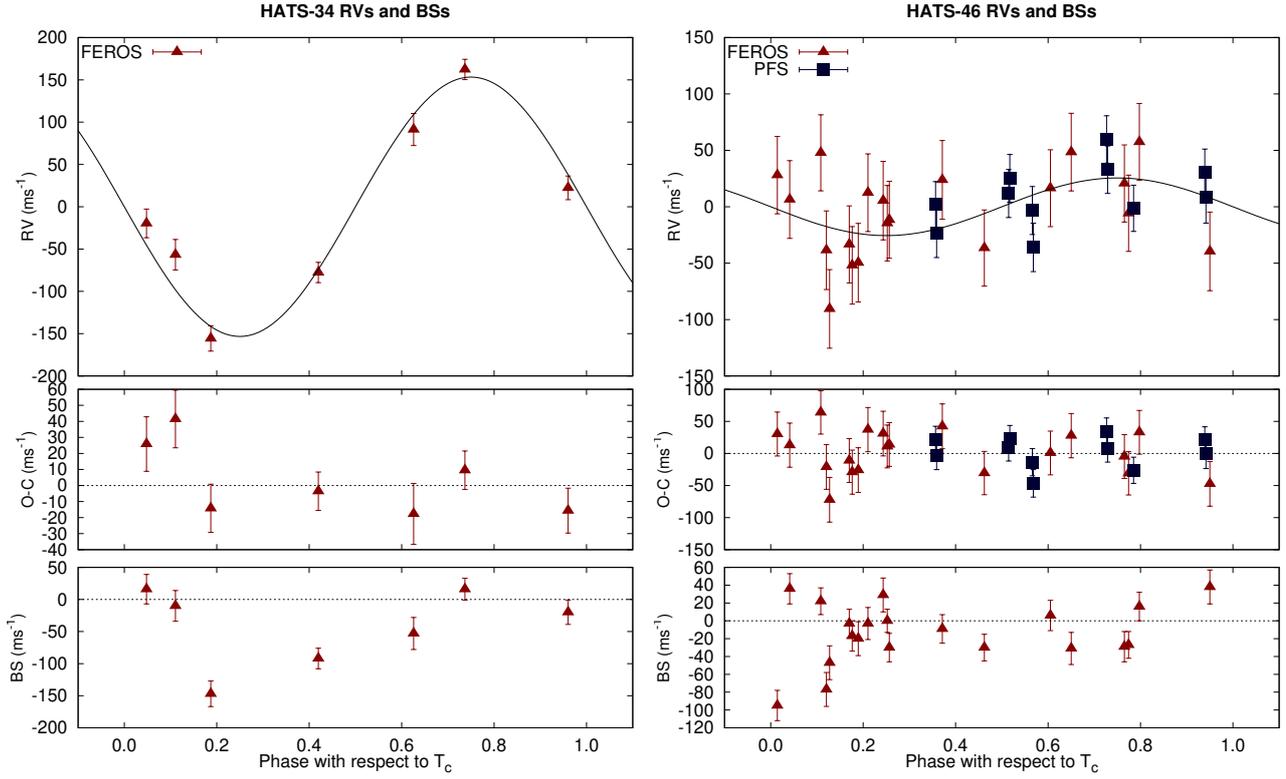

**Figure 5.** Radial Velocity Measurements Measurements of HATS-34 (*left*) and HATS-46 (*right*). We show observations phase-folded at the transit ephemeris. The instruments shown for each system are labelled in the corner. The *top* panel shows the measurements and best fit model (black line). The *middle* panel shows the velocity $O - C$ residuals from the best-fit model and the *bottom* panel shows the bisector spans.

**Table 2.** Summary of stellar and planetary parameters for HATS-34

| | | | | | | |
|---|---|---|---|---|---|---|
| | | **HATS-34** | | | | |
| | | $e \equiv 0$ | | | $e$ varied | |
| | MIST Isochrone [a] | PARSEC Isochrone | PARSEC Empirical | MIST Isochrone | PARSEC Isochrone | PARSEC Empirical |
| Stellar Parameters | | | | | | |
| $M_\star$ $(M_\odot)$ .. | $0.952^{+0.040}_{-0.020}$ | $0.953^{+0.050}_{-0.056}$ | $0.699^{+0.039}_{-0.083}$ | $0.970^{+0.040}_{-0.030}$ | $0.948^{+0.090}_{-0.041}$ | $0.726^{+0.156}_{-0.084}$ |
| $R_\star$ $(R_\odot)$ .... | $0.9381 \pm 0.0080$ | $0.9092 \pm 0.0084$ | $0.9029 \pm 0.0084$ | $0.9369 \pm 0.0082$ | $0.972^{+0.103}_{-0.039}$ | $0.9043 \pm 0.0093$ |
| $\log g_\star$ (cgs) . | $4.474 \pm 0.018$ | $4.497 \pm 0.022$ | $4.372 \pm 0.044$ | $4.481 \pm 0.020$ | $4.445 \pm 0.037$ | $4.386 \pm 0.068$ |
| $\rho_\star$ (g cm$^{-3}$) . | $1.634 \pm 0.078$ | $1.773 \pm 0.100$ | $1.340^{+0.056}_{-0.125}$ | $1.663 \pm 0.088$ | $1.47 \pm 0.19$ | $1.38^{+0.30}_{-0.13}$ |
| $L_\star$ $(L_\odot)$ .... | $0.702 \pm 0.012$ | $0.698 \pm 0.013$ | $0.715 \pm 0.012$ | $0.701 \pm 0.012$ | $0.80^{+0.21}_{-0.12}$ | $0.717 \pm 0.012$ |
| $T_{\rm eff\star}$ (K).... | $5456 \pm 19$ | $5543 \pm 20$ | $5586 \pm 25$ | $5456 \pm 19$ | $5530^{+220}_{-110}$ | $5590 \pm 24$ |
| [Fe/H] (dex) | $0.174^{+0.074}_{-0.029}$ | $0.109^{+0.041}_{-0.093}$ | $0.167 \pm 0.046$ | $0.210 \pm 0.051$ | $0.1558^{+0.0173}_{-0.0089}$ | $0.149 \pm 0.023$ |
| Age (Gyr) .. | $5.6^{+1.3}_{-2.2}$ | $4.3^{+1.5}_{-2.2}$ | ... | $4.7^{+1.8}_{-2.5}$ | $7.0^{+3.2}_{-4.3}$ | ... |
| $A_V$ (mag) .. | $0.063 \pm 0.013$ | $0.065^{+0.012}_{-0.016}$ | $0.057 \pm 0.013$ | $0.061 \pm 0.013$ | $0.063^{+0.011}_{-0.016}$ | $0.058 \pm 0.015$ |
| Distance (pc) | $517.2 \pm 3.7$ | $519.7 \pm 4.1$ | $517.0 \pm 3.4$ | $517.1 \pm 3.2$ | $518^{+28}_{-13}$ | $518.2 \pm 3.3$ |
| Selected Planetary Parameters | | | | | | |
| $M_P$ $(M_J)$ ... | $0.951 \pm 0.050$ | $0.927 \pm 0.059$ | $0.757 \pm 0.057$ | $0.957 \pm 0.052$ | $0.938 \pm 0.058$ | $0.794 \pm 0.088$ |
| $R_P$ $(R_J)$ ... | $1.282 \pm 0.064$ | $1.188^{+0.054}_{-0.070}$ | $1.222 \pm 0.054$ | $1.238^{+0.054}_{-0.083}$ | $1.327^{+0.103}_{-0.058}$ | $1.239^{+0.031}_{-0.042}$ |
| $\rho_P$ (g cm$^{-3}$) . | $0.558^{+0.115}_{-0.072}$ | $0.685^{+0.126}_{-0.077}$ | $0.513 \pm 0.080$ | $0.631^{+0.163}_{-0.091}$ | $0.492 \pm 0.088$ | $0.533 \pm 0.091$ |
| $\log g_P$ (cgs) . | $3.155^{+0.061}_{-0.045}$ | $3.211^{+0.051}_{-0.038}$ | $3.100 \pm 0.052$ | $3.194^{+0.069}_{-0.049}$ | $3.118 \pm 0.054$ | $3.119^{+0.046}_{-0.064}$ |
| $a$ (AU) ..... | $0.03165^{+0.00043}_{-0.00022}$ | $0.03165^{+0.00033}_{-0.00063}$ | $0.02854^{+0.00052}_{-0.00118}$ | $0.03184^{+0.00043}_{-0.00033}$ | $0.03160^{+0.00096}_{-0.00046}$ | $0.0289^{+0.0019}_{-0.0012}$ |
| $T_{\rm eq}$ (K) ..... | $1431.2 \pm 9.9$ | $1434 \pm 11$ | $1516^{+30}_{-15}$ | $1427 \pm 11$ | $1484^{+71}_{-50}$ | $1507^{+29}_{-47}$ |

[a] We suggest adopting these values

**Figure 6. Isochrones and Spectral Energy Distribution (SED) for HATS-34 (*left*) and HATS-46 (*right*).** The *top* panel shows the absolute $G$ magnitude vs. the de-reddened $BP - RP$ colour compared to the theoretical isochrones, indicated by the black lines, the stellar evolution tracks, indicated by the dotted green lines, from the MIST models interpolated at the best-fit metallicity of the host. The age of each isochrone is labelled in Gyrs in black and the mass of each evolution track is listed in $M_\odot$ in green. The filled blue circles show the measured reddening and distance-corrected values from *Gaia* DR2 and the blue lines indicate the $1\sigma$ and $2\sigma$ confidence regions, including the estimated systematic errors in the photometry. The *middle* panel shows the SED as measured through the filters indicated on the figure. The magnitudes are plotted without correction for distance or extinction. 200 model SEDs randomly selected from the MCMC posterior distribution are overlaid on the plot. The *bottom* panel shows the $O - C$ residuals from the best-fit model SED.

**Table 3. Summary of stellar and planetary parameters for HATS-46**

| | HATS-46 | | | | | |
|---|---|---|---|---|---|---|
| | | $e \equiv 0$ | | | $e$ varied | |
| | MIST Isochrone [a] | PARSEC Isochrone | PARSEC Empirical | MIST Isochrone | PARSEC Isochrone | PARSEC Empirical |
| **Stellar Parameters** | | | | | | |
| $M_\star$ ($M_\odot$) .. | $0.869 \pm 0.023$ | $0.820^{+0.018}_{-0.012}$ | $1.021^{+0.059}_{-0.103}$ | $0.867 \pm 0.021$ | $0.820^{+0.018}_{-0.013}$ | $0.46^{+0.39}_{-0.22}$ |
| $R_\star$ ($R_\odot$) .... | $0.894 \pm 0.010$ | $0.8649 \pm 0.0068$ | $0.868 \pm 0.011$ | $0.8933 \pm 0.0084$ | $0.8621 \pm 0.0072$ | $0.869 \pm 0.011$ |
| $\log g_\star$ (cgs). | $4.474 \pm 0.019$ | $4.478 \pm 0.012$ | $4.571 \pm 0.039$ | $4.474 \pm 0.016$ | $4.481 \pm 0.013$ | $4.23 \pm 0.25$ |
| $\rho_\star$ (g cm$^{-3}$) | $1.715 \pm 0.097$ | $1.786^{+0.069}_{-0.053}$ | $2.204^{+0.096}_{-0.170}$ | $1.715 \pm 0.078$ | $1.803 \pm 0.067$ | $1.00^{+0.82}_{-0.56}$ |
| $L_\star$ ($L_\odot$) ... | $0.633 \pm 0.012$ | $0.623 \pm 0.011$ | $0.643 \pm 0.012$ | $0.631 \pm 0.010$ | $0.620 \pm 0.011$ | $0.644 \pm 0.012$ |
| $T_{\rm eff\star}$ (K) .... | $5451 \pm 19$ | $5523 \pm 14$ | $5550 \pm 28$ | $5446 \pm 18$ | $5524 \pm 16$ | $5545 \pm 29$ |
| [Fe/H] (dex) | $-0.029 \pm 0.039$ | $-0.172^{+0.030}_{-0.022}$ | $-0.125 \pm 0.045$ | $-0.032 \pm 0.041$ | $-0.175 \pm 0.028$ | $-0.099 \pm 0.042$ |
| Age (Gyr) .. | $8.4 \pm 1.9$ | $10.8 \pm 1.4$ | $\cdots$ | $8.6 \pm 1.7$ | $10.6 \pm 1.5$ | $\cdots$ |
| $A_V$ (mag) .. | $0.043 \pm 0.012$ | $0.043 \pm 0.010$ | $0.037 \pm 0.011$ | $0.043 \pm 0.011$ | $0.044 \pm 0.012$ | $0.036 \pm 0.010$ |
| Distance (pc) | $458.8 \pm 4.0$ | $458.8 \pm 3.5$ | $459.0 \pm 3.5$ | $457.9 \pm 3.3$ | $457.4 \pm 3.5$ | $459.6 \pm 3.8$ |
| | | | | | | |
| **Selected Planetary Parameters** | | | | | | |
| $M_P$ ($M_J$) ... | $0.158 \pm 0.042$ | $0.149 \pm 0.048$ | $0.145 \pm 0.040$ | $0.164 \pm 0.035$ | $0.147 \pm 0.046$ | $0.098 \pm 0.049$ |
| $R_P$ ($R_J$) .... | $0.951 \pm 0.029$ | $0.919 \pm 0.019$ | $0.905 \pm 0.022$ | $0.940 \pm 0.022$ | $0.912 \pm 0.017$ | $0.909^{+0.017}_{-0.024}$ |
| $\rho_P$ (g cm$^{-3}$) | $0.227 \pm 0.065$ | $0.238 \pm 0.079$ | $0.243 \pm 0.074$ | $0.244 \pm 0.058$ | $0.240 \pm 0.078$ | $0.163 \pm 0.084$ |
| $\log g_P$ (cgs) . | $2.64 \pm 0.14$ | $2.64^{+0.12}_{-0.17}$ | $2.64 \pm 0.13$ | $2.663^{+0.083}_{-0.114}$ | $2.64^{+0.12}_{-0.15}$ | $2.47^{+0.19}_{-0.26}$ |
| $a$ (AU) ..... | $0.05272 \pm 0.00045$ | $0.05170^{+0.00037}_{-0.00026}$ | $0.0556^{+0.0011}_{-0.0019}$ | $0.05267 \pm 0.00042$ | $0.05171 \pm 0.00035$ | $0.0427 \pm 0.0080$ |
| $T_{\rm eq}$ (K) ..... | $1082.1 \pm 8.2$ | $1089.1 \pm 5.8$ | $1057.0^{+18.2}_{-9.6}$ | $1083.1 \pm 7.3$ | $1088.0 \pm 5.8$ | $1220^{+170}_{-130}$ |

[a] We suggest adopting these values

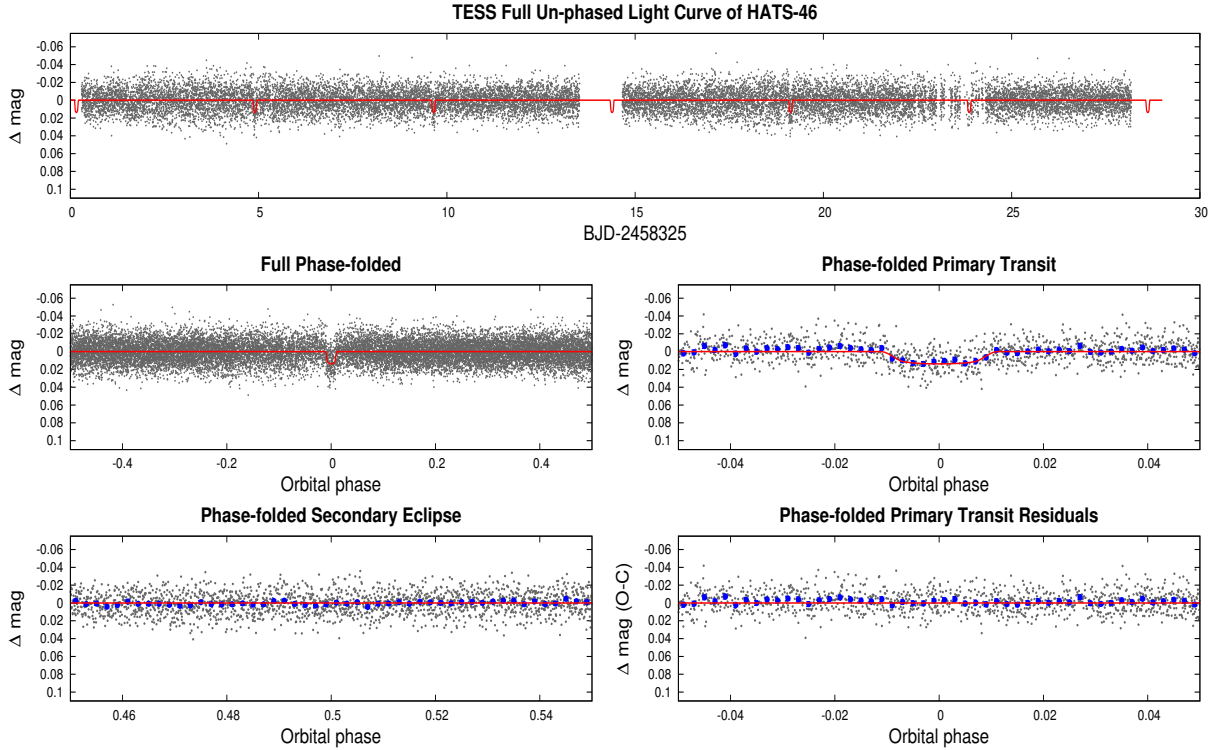

**Figure 7. TESS Light Curve for HATS-46.** The *top* plot shows the full light curve (with the model in red). The *middle* row of plots shows the full light curve after it has been phase folded (model in red) and a zoomed in plot of the primary transit with the model (red line) and the light curve binned in phase with a bin size of 0.002 (blue dots) overlaid. The *bottom* row shows the phase folded light curve zoomed in on the secondary eclipse with the model in red and the phase-folded residuals zoomed in on the primary transit.

limited by systematic uncertainties in the stellar evolution models. We suggest adopting the circular-orbit MIST isochrone-based parameters for both systems, which yields a stellar mass and radius of $0.952^{+0.040}_{-0.020}M_\odot$ and $0.9381 \pm 0.0080R_\odot$ for HATS-34, and $0.869 \pm 0.023M_\odot$ and $0.894 \pm 0.010R_\odot$ for HATS-46, and a planetary mass and radius of $0.951 \pm 0.050M_J$ and $1.282 \pm 0.064R_J$ for HATS-34b, and $0.158 \pm 0.042M_J$ and $0.951 \pm 0.029R_J$ for HATS-46b.

The re-analysis of these systems differs from prior work on these objects by incorporating the *Gaia* DR2 parallax and photometry measurements directly into a global fit of all the relevant photometric, spectroscopic, and astrometric observations of each system, by imposing theoretical isochrone constraints using the MIST models, and by making use of the *TESS* light curves for the first time. Below we compare our results to the results from prior studies.

### 5.2 Comparison with Values From Discovery Papers.

We compare the published values and the values produced by isochrone modeling in Table 4 for both systems. The parameters that we compare include the change in uncertainty of planetary mass, planetary radius, semi-major axis, equilibrium temp, stellar mass, stellar radius, effective temperature, metallicity, and age.

For HATS-34 the isochrone model produced uncertainties that were less by factors of 0.69 to 0.17 for the stellar and planetary masses and radii. The greatest decrease in uncertainty, 0.17, was for the stellar radius, as expected due to the inclusion of *Gaia* DR2 data. The uncertainty in stellar mass increased by a factor of 1.45.

The uncertainty in planetary mass and radius improved by factors of 0.69 and 0.33, respectively.

For HATS-46 the isochrone model produced lower certainties by factors of 0.85 to 0.19. The isochrone constraint tightened the stellar radius with a decrease in uncertainty of a factor of 0.19, as expected given the inclusion of *Gaia* DR2 data. The uncertainty in stellar mass decreased by a factor of 0.85. The uncertainty in planetary mass and radius improved by factors of 0.67 and 0.4, respectively.

### 5.3 Comparison With Previously Updated Values

We also compare the values found for HATS-34b and HATS-46b to the updated values published by Johns et al. (2018). Their results did not produce improvements in the mass or radius precision for either HATS-34b or HATS-46b. Their published values are found in Table 4.

In particular, for HATS-34 they find almost no improvement in $R_p$ over the discovery values, because in this case the degeneracy between the impact parameter $b$ and $R_p/R_\star$ for the grazing system results in a poor constraint on the published $R_p/R_\star$ and hence on $R_p$. Our work allows the improved constraint on the stellar density to also tighten the constraint on $R_p/R_\star$ and hence on $R_p$ as well.

### 5.4 Comparison with Analytically Predicted Uncertainties Using *Gaia*

Stevens et al. (2018) derived analytical estimates of uncertainties for stellar mass and radii in the current era where *TESS* and *Gaia*

**Table 4.** Comparison with Previously Published Parameters for HATS-34 and HATS-46.

| | **HATS-34** | | | **HATS-46** | | |
|---|---|---|---|---|---|---|
| Parameter | Discovery Paper | Johns et. al | This Work | Discovery Paper | Johns et. al | This Work |
| Stellar Parameters | | | | | | |
| $M_\star (M_\odot)$ | $0.955 \pm 0.031$ | $\cdots$ | $0.952^{+0.040}_{-0.020}$ | $0.917 \pm 0.027$ | $\cdots$ | $0.869 \pm 0.023$ |
| $R_\star (R_\odot)$ | $0.980 \pm 0.047$ | $0.962^{+0.020}_{-0.024}$ | $0.9381 \pm 0.0080$ | $0.853^{+0.040}_{-0.030}$ | $0.896^{+0.021}_{-0.026}$ | $0.894 \pm 0.010$ |
| $T_{eff}$ (K) | $5380 \pm 73$ | $\cdots$ | $5456 \pm 19$ | $5495 \pm 69$ | $\cdots$ | $5451 \pm 19$ |
| Metallicity | $0.250 \pm 0.070$ | $\cdots$ | $0.174^{+0.074}_{-0.029}$ | $-0.060 \pm 0.046$ | $\cdots$ | $-0.029 \pm 0.039$ |
| Age (Gyr) | $7.7 \pm 2.7$ | $\cdots$ | $5.6^{+1.5}_{-2.2}$ | $3^{+3.4}_{-2}$ | $\cdots$ | $8.4 \pm 1.9$ |
| $L_\star (L_\odot)$ | $0.724 \pm 0.089$ | $0.695 \pm 0.016$ | $0.695 \pm 0.015$ | $0.589 \pm 0.070$ | $0.635 \pm 0.015$ | $0.6173 \pm 0.0099$ |
| Planetary Parameters | | | | | | |
| $M_P (M_J)$ | $0.941 \pm 0.072$ | $0.941 \pm 0.072$ | $0.951 \pm 0.050$ | $0.173 \pm 0.062$ | $0.173 \pm 0.062$ | $0.158 \pm 0.042$ |
| $R_P (R_J)$ | $1.43 \pm 0.19$ | $1.405 \pm 0.275$ | $1.282 \pm 0.064$ | $0.903^{+0.058}_{-0.045}$ | $0.954^{+0.098}_{-0.084}$ | $0.951 \pm 0.029$ |
| a (AU) | $0.03166 \pm 0.00034$ | $\cdots$ | $0.03165^{+0.00043}_{-0.00022}$ | $0.05367 \pm 0.00053$ | $\cdots$ | $0.05272 \pm 0.00045$ |
| $T_{eq}$ (K) | $1445 \pm 42$ | $\cdots$ | $1431.2 \pm 9.9$ | $1054 \pm 29$ | $\cdots$ | $1082.1 \pm 8.2$ |

data are available. We compare our uncertainties in the PARSEC Empirical results (e=0) in this work to the analytic estimates based upon Equations 50-53 in their paper. We also compare the uncertainties of the MIST Circular model to their analytic predictions for isochrone based models using Equations 69-71 of their paper. The comparisons are reported in Table 5. We can see that for $\sigma_{M_\star}$ the analytical uncertainties are of the same order of magnitude as the ones found in this work with the exception of the PARSEC Empirical value for HATS-46. The same is true for $\sigma_{M_P}$, which show remarkably close uncertainties. $\sigma_{R_\star}$ is better constrained by this work for HATS-34 and HATS-46. $\sigma_{R_P}$ is of the same order of magnitude with the exception of the empirical comparison for HATS-34.

These discrepancies and tighter constraints on our derived parameters are likely due to a few parts of the Stevens et al. (2018) method for estimating the isochrone-based uncertainties that could lead to overestimates, compared to what we found.

First, Stevens et al. (2018) effectively treat the isochrones as power-law mass–radius relations whose coefficients depend on $\frac{Fe}{H}$, $T_{eff}$ and $\log_g$. These power-law relations, however, do not constrain the set of $\frac{Fe}{H}$, $T_{eff}$ and $\log_g$ combinations to a physically permitted range. The isochrone models, however, impose a strong constraint on the set of allowed combinations, especially if looking at sub-Solar-mass stars, and if assuming an upper limit on the allowed ages of stars. Indeed some of the parameter combinations in our MCMC analyses do not match to stellar evolution models that are younger than the age of the Galaxy, and this causes those combinations to be rejected in the MCMC chain. This constraint then leads to smaller scatter on the output stellar masses and radii than what would be estimated from simply perturbing the power-law relations, as done to get Equations 69 through 71.

Furthermore, Equations 69 through 71 of Stevens et al. (2018) do not take into account the possibility that the stellar mass and radius can be over-determined by simultaneously fitting the parallax/bolometric flux, the effective temperature, the metallicity, and the transit-density, while simultaneously imposing the stellar isochrone constraints. Equations 69 through 71 of Stevens et al. (2018) neglect the parallax and bolometric flux uncertainties, and the possibility that these can put an even tighter constraint on the stellar radius and mass, when coupled with the other observables, than can be done with only the transit-inferred density, $\frac{Fe}{H}$, and $T_{eff}$.

Additionally, in computing the analytic estimates we are assuming a 0.02 mag uncertainty for the bolometric flux, however this yields an overestimate on the effective bolometric flux uncertainty since we are actually fitting multiple photometric observations, each with an uncertainty of 0.02 mag. On the other hand, for HATS-34b, the grazing transits lead to a larger uncertainty on the density $\rho_\star$ than what one would estimate from the uncertainty on the ingress/egress duration $T_{12}$, which the Stevens et al. (2018) formulae assume sets the $\rho_\star$ uncertainty. Both of these factors can explain the discrepancies between the analytic estimates for the empirical uncertainties, and the actual uncertainties that we find in this paper.

### 5.5 Systematic Uncertainties in Stellar Parameters

It is important to note that the isochrone uncertainties do not account for systematic uncertainties in the stellar evolution models themselves, which may actually now be the dominant source of uncertainty in characterising planetary systems such as these. It is difficult to provide reliable quantitative estimates for these uncertainties, however the differences in estimated stellar radii between the MIST and PARSEC models suggest that there is at least a 3 per cent systematic error in this parameter from the stellar models. Based on comparisons between stellar evolution models and observations of detached eclipsing binary systems, the models appear to reproduce the measured masses and radii of solar-type stars to within ∼ 1 – 3 per cent precision (e.g., Torres et al. 2010), which is comparable to the 3 per cent estimated systematic uncertainty on the stellar radii.

Furthermore, we recognise that there are additional factors that could increase the uncertainty in the stellar parameters. Some additional possible sources that could increase uncertainty include star spots and plages, incorrect zero-points and filter transmission profiles, systematic errors that may result from assuming a quadratic limb-darkening law, and associated errors in the theoretical limb darkening coefficients used to place prior constraints on the fitted parameters, and errors in the assumed extinction law or the Galactic dust model used in placing a prior on the extinction. By using two different sets of isochrones we add a point of comparison in the uncertainties in the stellar parameters for each of the systems, but further work is needed to assess the impact of these other possible systematic error sources.

**Table 5.** Comparison with Analytically Predicted Uncertainties for HATS-34 and HATS-46.

| Uncertainty | HATS-34 | | | | HATS-46 | | | |
|---|---|---|---|---|---|---|---|---|
| | Analytical Empirical | PARSEC Empirical | Analytical Isochrone | MIST Circular | Analytical Empirical | PARSEC Empirical | Analytical Isochrone | MIST Circular |
| $\sigma_{M_\star}$ ($M_\odot$) | 0.046 | 0.092 | 0.087 | 0.045 | 0.087 | 0.120 | 0.018 | 0.023 |
| $\sigma_{R_\star}$ ($R_\odot$) | 0.013 | 0.009 | 0.030 | 0.008 | 0.013 | 0.011 | 0.031 | 0.010 |
| $\sigma_{M_P}$ ($M_J$) | 0.054 | 0.057 | 0.073 | 0.050 | 0.043 | 0.040 | 0.041 | 0.042 |
| $\sigma_{R_P}$ ($R_J$) | 0.124 | 0.054 | 0.040 | 0.064 | 0.047 | 0.042 | 0.033 | 0.029 |

### 5.6 Conclusions and Further Work

HATS-34 and HATS-46 were chosen to reanalyse because they were in the first sector observed by *TESS*. There are approximately 300 more planets that have been observed by *TESS* and have yet to be re-analysed. Approximately 200 of these are ground-based discoveries. The combination of *TESS* and *Gaia* data into the re-characterisation of these systems is valuable in tightening constraints on the parameters. For example, if instead of performing a new fit the stellar radii found by *Gaia* were used with previously determined $R_P/R_\star$ values to determine the planetary radii, the tighter constraint on orbital inclination allowed for by re-analysis would be neglected and thus the planetary radii would not be as precisely found. This is particularly important for the grazing system HATS-34b. In this case if we simply multiply the fitted $R_\star$ value with the $R_P/R_\star$ value taken from the discovery paper, we find an uncertainty on $R_P$ that is a factor of two greater than what we found through our full re-analysis.

While we find that our re-analysis has improved the precision of most parameters, the precision is not improved for every parameter. As seen in Table 4, for HATS-34 the stellar mass in the discovery paper has an uncertainty that is 0.7 times less than what we find through our re-analysis.

Nonetheless, by incorporating the parallax from Gaia into the analysis, the accuracy of all of the derived parameters should be improved compared to the discovery paper. An example of this is the stellar mass of HATS-46. While the uncertainty on this parameter is only a modest 1.17 times greater in the discovery paper than in this paper, the new stellar mass estimate lies below the error range from the prior work, illustrating that the pre-*Gaia* analyses are not just less precise than what is possible with *Gaia*, they may also be inaccurate. This shows that the re-analysis is preferable to simply adopting the prior fits and adjusting the stellar parameters.

In summary, this paper presents revised characterisation of systems HATS-34 and HATS-46 making use of *Gaia* high-precision parallax and photometry. Previous work had higher uncertainties due to the lack of the high-precision parallax and photometry. We have demonstrated that pre-*Gaia* analyses may be both imprecise and inaccurate leading to the need to reanalyse systems found prior to *Gaia*.

### DATA AVAILABILITY

The observational data underlying this article are available from various publications and data repositories as referenced in the article. Codes used to analyse the data will be made available upon reasonable request to the authors.

### ACKNOWLEDGEMENTS

We thank the anonymous referees for their thoughtful and helpful feedback to improve the quality of the paper. We thank W. Bhatti and G. Bakos for help in managing the computer systems used in carrying out this work. J.H. acknowledges support from NASA grants NNX17AB61G and 80NSSC19K0386. This work has made use of data from the European Space Agency (ESA) mission *Gaia* (`https://www.cosmos.esa.int/gaia`), processed by the *Gaia* Data Processing and Analysis Consortium (DPAC, `https://www.cosmos.esa.int/web/gaia/dpac/consortium`). Funding for the DPAC has been provided by national institutions, in particular the institutions participating in the *Gaia* Multilateral Agreement. This research has made use of the NASA Exoplanet Archive, which is operated by the California Institute of Technology, under contract with the National Aeronautics and Space Administration under the Exoplanet Exploration Program.

## APPENDIX A: CORNER PLOTS

Figures A1 and A2 show corner plots of the parameters that were varied in our DEMCMC analysis of HATS-34 and HATS-46, respectively. Parameters that show strong correlations are cases of well-known degeneracies.

This paper has been typeset from a TEX/LATEX file prepared by the author.

**Table 6.** Orbital parameters for HATS-34b

| | **HATS-34** | | | | | |
| | **e ≡ 0** | | | **e varied** | | |
| | MIST Isochrone[a] | PARSEC Isochrone | PARSEC Empirical | MIST Isochrone | PARSEC Isochrone | PARSEC Empirical |
|---|---|---|---|---|---|---|
| $P$ (days) | $2.1061619 \pm 0.0000013$ | $2.1061617 \pm 0.0000013$ | $2.1061624 \pm 0.0000015$ | $2.1061624 \pm 0.0000014$ | $2.1061625 \pm 0.0000013$ | $2.1061627 \pm 0.0000013$ |
| $T_c$ (BJD)[a] | $2457346.74007 \pm 0.00046$ | $2457327.78472 \pm 0.00050$ | $2457355.16458 \pm 0.00049$ | $2457344.63402 \pm 0.00053$ | $2457454.15435 \pm 0.00054$ | $2457416.24331 \pm 0.00058$ |
| $T_{14}$ (days)[a] | $0.0622 \pm 0.0013$ | $0.0610 \pm 0.0016$ | $0.0643 \pm 0.0018$ | $0.0631 \pm 0.0015$ | $0.0645 \pm 0.0018$ | $0.0648 \pm 0.0021$ |
| $T_{12} = T_{34}$ (days)[a] | $0.0772 \pm 0.0013$ | $0.0754 \pm 0.0024$ | $0.0802 \pm 0.0024$ | $0.0780 \pm 0.0035$ | $0.0803 \pm 0.0024$ | $0.0808 \pm 0.0026$ |
| $a/R_\star$ | $7.27 \pm 0.11$ | $7.47 \pm 0.14$ | $6.800^{+0.094}_{-0.37}$ | $7.31 \pm 0.13$ | $7.02 \pm 0.31$ | $6.87^{+0.46}_{-0.22}$ |
| $\zeta/R_\star$[b] | $55.0^{+3.0}_{-2.3}$ | $53.2 \pm 3.1$ | $54.1^{+6.4}_{-4.6}$ | $52.1 \pm 3.5$ | $54.8^{+2.7}_{-3.6}$ | $53.9 \pm 3.6$ |
| $R_p/R_\star$ | $0.1402 \pm 0.0067$ | $0.1342 \pm 0.0065$ | $0.1389 \pm 0.0057$ | $0.1356 \pm 0.0072$ | $0.1405 \pm 0.0041$ | $0.1408 \pm 0.0056$ |
| $b^2$ | $0.845^{+0.016}_{-0.014}$ | $0.824^{+0.019}_{-0.024}$ | $0.853^{+0.033}_{-0.031}$ | $0.830^{+0.024}_{-0.029}$ | $0.855^{+0.020}_{-0.023}$ | $0.853^{+0.021}_{-0.023}$ |
| $b \equiv a \cos i/R_\star$ | $0.9193^{+0.0085}_{-0.0079}$ | $0.908^{+0.010}_{-0.013}$ | $0.923^{+0.018}_{-0.017}$ | $0.911^{+0.013}_{-0.016}$ | $0.925^{+0.011}_{-0.012}$ | $0.923^{+0.011}_{-0.012}$ |
| $i$ (deg) | $82.73 \pm 0.14$ | $83.04^{+0.12}_{-0.18}$ | $82.19^{+0.21}_{-0.38}$ | $82.96 \pm 0.27$ | $82.47 \pm 0.55$ | $82.42^{+0.75}_{-0.44}$ |
| **HATSouth dilution factors[c]** | | | | | | |
| HS Dilution factor | $0.872 \pm 0.084$ | $0.864 \pm 0.081$ | $0.871 \pm 0.088$ | $0.862 \pm 0.085$ | $0.884 \pm 0.082$ | $0.882 \pm 0.083$ |
| **Limb-darkening coefficients[d]** | | | | | | |
| $c_1, r$ | $0.39 \pm 0.17$ | $0.36 \pm 0.15$ | $0.35 \pm 0.15$ | $0.36 \pm 0.17$ | $0.39 \pm 0.15$ | $0.37 \pm 0.13$ |
| $c_2, r$ | $0.41^{+0.14}_{-0.20}$ | $0.38 \pm 0.16$ | $0.40 \pm 0.15$ | $0.37 \pm 0.16$ | $0.41^{+0.10}_{-0.17}$ | $0.44 \pm 0.15$ |
| $c_1, R$ | $0.33 \pm 0.13$ | $0.33 \pm 0.13$ | $0.30 \pm 0.14$ | $0.32 \pm 0.14$ | $0.30 \pm 0.13$ | $0.30 \pm 0.15$ |
| $c_2, R$ | $0.29 \pm 0.17$ | $0.26 \pm 0.17$ | $0.27 \pm 0.17$ | $0.27 \pm 0.16$ | $0.30^{+0.16}_{-0.12}$ | $0.30 \pm 0.16$ |
| $c_1, T$ | $0.39 \pm 0.14$ | $0.34 \pm 0.14$ | $0.33 \pm 0.15$ | $0.35 \pm 0.14$ | $0.36 \pm 0.15$ | $0.36 \pm 0.13$ |
| $c_2, T$ | $0.32 \pm 0.15$ | $0.38 \pm 0.15$ | $0.36 \pm 0.17$ | $0.36^{+0.14}_{-0.19}$ | $0.36 \pm 0.15$ | $0.35 \pm 0.18$ |
| **RV parameters** | | | | | | |
| $K$ (m s$^{-1}$) | $154.6 \pm 7.3$ | $152.2 \pm 8.3$ | $152.0 \pm 7.2$ | $154.2 \pm 7.6$ | $151.9 \pm 7.3$ | $153.2 \pm 7.1$ |
| $e$[f] | ⋯ | ⋯ | ⋯ | $< 0.075$ | $< 0.096$ | $< 0.097$ |
| RV jitter FEROS (m s$^{-1}$)[g] | $< 10.5$ | $< 3.6$ | $< 9.3$ | $< 0.4$ | $< 0.6$ | $< 7.4$ |

[a] Times are in Barycentric Julian Date. $T_c$: Reference epoch of mid transit that minimises the correlation with the orbital period. $T_{12}$: total transit duration, time between first to last contact; $T_{12} = T_{34}$: ingress/egress time, time between first and second, or third and fourth contact.

[b] Reciprocal of the half duration of the transit used as a jump parameter in our MCMC analysis in place of $a/R_\star$. It is related to $a/R_\star$ by the expression $\zeta/R_\star = a/R_\star (2\pi(1 + e \sin \omega))/(P\sqrt{1 - b^2}\sqrt{1 - e^2})$ (Bakos et al. 2010).

[c] Scaling factor applied to the model transit that is fit to the HATSouth light curves. This factor accounts for dilution of the transit due to blending from neighbouring stars and over-filtering of the light curve. These factors are varied in the fit, with independent values adopted for each light curve.

[d] Values for a quadratic law. For the isochrone and empirical models these are allowed to vary in the fit, with prior constraints adopted from the tabulations by Claret et al. (2012, 2013); Claret (2018). For the HPS models, the tabulated limb darkening coefficients are adopted according to the stellar atmospheric parameters for each link in the Markov Chain, but the assumed values were not output by the modelling software, so we do not list them here.

[f] 95% confidence upper limit.

[g] Term added in quadrature to the formal RV uncertainties for each instrument. This is treated as a free parameter in the fitting routine. In cases where the jitter is consistent with zero, we list its 95 per cent confidence upper limit.

[h] Correlation coefficient between the planetary mass $M_p$ and radius $R_p$, estimated from the posterior parameter distribution. This was not calculated for the HPS model, where the planetary mass was not fit directly in the same Markov Chain as the planetary radius.

[i] The Safronov number is given by $\Theta = \frac{1}{2}(V_{esc}/V_{orb})^2 = (a/R_p)(M_p/M_\star)$ (see Hansen & Barman 2007).

[j] Incoming flux per unit surface area, averaged over the orbit.

[k] We suggest adopting these values.

**Table 7.** Orbital parameters for HATS-46b

| | | $e \equiv 0$ | | | $e$ varied | |
|---|---|---|---|---|---|---|
| **HATS-46** | MIST Isochrone [a] | PARSEC Isochrone | PARSEC Empirical | MIST Isochrone | PARSEC Isochrone | PARSEC Empirical |
| $P$ (days) .......... | $4.7423749 \pm 0.0000043$ | $4.7423733 \pm 0.0000046$ | $4.7423743 \pm 0.0000044$ | $4.7423739 \pm 0.0000042$ | $4.7423754 \pm 0.0000040$ | $4.7423732 \pm 0.0000039$ |
| $T_c$ (BJD) [a] .......... | $2457149.05166 \pm 0.00056$ | $2457300.80755 \pm 0.00055$ | $2457352.97349 \pm 0.00052$ | $2457186.99049 \pm 0.00058$ | $2457239.15657 \pm 0.00053$ | $2457210.70220 \pm 0.00060$ |
| $T_{14}$ (days) [a] .......... | $0.1036 \pm 0.0016$ | $0.1028 \pm 0.0013$ | $0.1013 \pm 0.0016$ | $0.1015 \pm 0.0019$ | $0.1017 \pm 0.0015$ | $0.1000 \pm 0.0025$ |
| $T_{12} = T_{34}$ (days) [a] ...... | $0.0182 \pm 0.0011$ | $0.01763 \pm 0.00067$ | $0.0149 \pm 0.0012$ | $0.0161 \pm 0.0014$ | $0.01624 \pm 0.00096$ | $0.0149 \pm 0.0019$ |
| $a/R_\star$ .......... | $12.68 \pm 0.23$ | $12.86 \pm 0.15$ | $13.79^{+0.20}_{-0.37}$ | $12.68 \pm 0.19$ | $12.90 \pm 0.16$ | $10.6 \pm 2.0$ |
| $\zeta/R_\star$ [b] .......... | | | $23.06 \pm 0.43$ | | | $23.35 \pm 0.41$ |
| $R_p/R_\star$ .......... | $0.1093 \pm 0.0026$ | $0.1092 \pm 0.0019$ | $0.1072 \pm 0.0023$ | $0.1082 \pm 0.0021$ | $0.1087 \pm 0.0015$ | $0.1076 \pm 0.0025$ |
| $b^2$ .......... | $0.474^{+0.026}_{-0.036}$ | $0.462^{+0.023}_{-0.025}$ | $0.374^{+0.043}_{-0.057}$ | $0.418^{+0.042}_{-0.045}$ | $0.417^{+0.035}_{-0.039}$ | $0.376^{+0.099}_{-0.082}$ |
| $b \equiv a\cos i/R_\star$ .......... | $0.688^{+0.026}_{-0.036}$ | $0.679^{+0.017}_{-0.019}$ | $0.611^{+0.034}_{-0.024}$ | $0.647^{+0.030}_{-0.033}$ | $0.645^{+0.027}_{-0.031}$ | $0.613^{+0.076}_{-0.071}$ |
| $i$ (deg) .......... | $86.89 \pm 0.16$ | $86.97 \pm 0.10$ | $87.46^{+0.12}_{-0.20}$ | $86.90 \pm 0.13$ | $87.03 \pm 0.11$ | $85.5^{+1.6}_{-3.2}$ |
| **HATSouth dilution factors** [c] | | | | | | |
| HS Dilution factor ....... | | | $0.882 \pm 0.056$ | | | $0.876 \pm 0.056$ |
| **Limb-darkening coefficients** [d] | | | | | | |
| $c_1, r$ .......... | $0.37 \pm 0.14$ | | $0.42 \pm 0.15$ | $0.50 \pm 0.15$ | $0.46 \pm 0.14$  $0.46 \pm 0.14$ | $0.47 \pm 0.14$ |
| $c_2, r$ .......... | $0.45 \pm 0.14$ | $0.40 \pm 0.16$ | $0.36 \pm 0.14$ | $0.36 \pm 0.15$ | $0.38 \pm 0.14$ | $0.37 \pm 0.14$ |
| $c_1, R$ .......... | $0.4887$ | $0.4887$ | | | $0.4887$ | |
| $c_2, R$ .......... | $0.1741$ | $0.1741$ | | | $0.1741$ | |
| $c_1, T$ .......... | $0.33 \pm 0.15$ | $0.32 \pm 0.14$ | $0.33 \pm 0.15$ | $0.36 \pm 0.16$ | | $0.31 \pm 0.15$ |
| $0.34 \pm 0.14$ | | | | | | |
| $c_2, T$ .......... | $0.33 \pm 0.17$ | $0.30 \pm 0.17$ | $0.29 \pm 0.16$ | $0.26 \pm 0.16$ | $0.30 \pm 0.17$ | $0.25 \pm 0.15$ |
| **RV parameters** | | | | | | |
| $K$ (m s$^{-1}$) .......... | $21.0 \pm 5.5$ | $20.6 \pm 6.6$ | $17.5 \pm 4.8$ | $22.1 \pm 4.8$ | $20.4 \pm 6.4$ | $21.6 \pm 7.7$ |
| $e$ [f] .......... | $\cdots$ | $\cdots$ | $\cdots$ | $< 0.207$ | $< 0.099$   $< 0.599$ | |
| RV jitter FEROS (m s$^{-1}$) [g] | | | | | | |

[a] Times are in Barycentric Julian Date. $T_c$: Reference epoch of mid transit that minimises the correlation with the orbital period. $T_{12}$: total transit duration, time between first to last contact; $T_{12} = T_{34}$: ingress/egress time, time between first and second, or third and fourth contact.

[b] Reciprocal of the half duration of the transit used as a jump parameter in our MCMC analysis in place of $a/R_\star$. It is related to $a/R_\star$ by the expression $\zeta/R_\star = a/R_\star(2\pi(1 + e\sin\omega))/(P\sqrt{1-b^2}\sqrt{1-e^2})$ (Bakos et al. 2010).

[c] Scaling factor applied to the model transit that is fit to the HATSouth light curves. This factor accounts for dilution of the transit due to blending from neighbouring stars and over-filtering of the light curve. These factors are varied in the fit, with independent values adopted for each light curve.

[d] Values for a quadratic law. For the isochrone and empirical models these are allowed to vary in the fit, with prior constraints adopted from the tabulations by Claret et al. (2012, 2013); Claret (2018). For the HPS models, the tabulated limb darkening coefficients are adopted according to the stellar atmospheric parameters for each link in the Markov Chain, but the assumed values were not output by the modelling software, so we do not list them here.

[f] 95% confidence upper limit.

[g] Term added in quadrature to the formal RV uncertainties for each instrument. This is treated as a free parameter in the fitting routine. In cases where the jitter is consistent with zero, we list its 95 per cent confidence upper limit.

[h] Correlation coefficient between the planetary mass $M_p$ and radius $R_p$ estimated from the posterior parameter distribution. This was not calculated for the HPS model, where the planetary mass was not fit directly in the same Markov Chain as the planetary radius.

[i] The Safronov number is given by $\Theta = \frac{1}{2}(V_{esc}/V_{orb})^2 = (a/R_p)(M_p/M_\star)$ (see Hansen & Barman 2007).

[j] Incoming flux per unit surface area, averaged over the orbit.

[k] We suggest adopting these values.

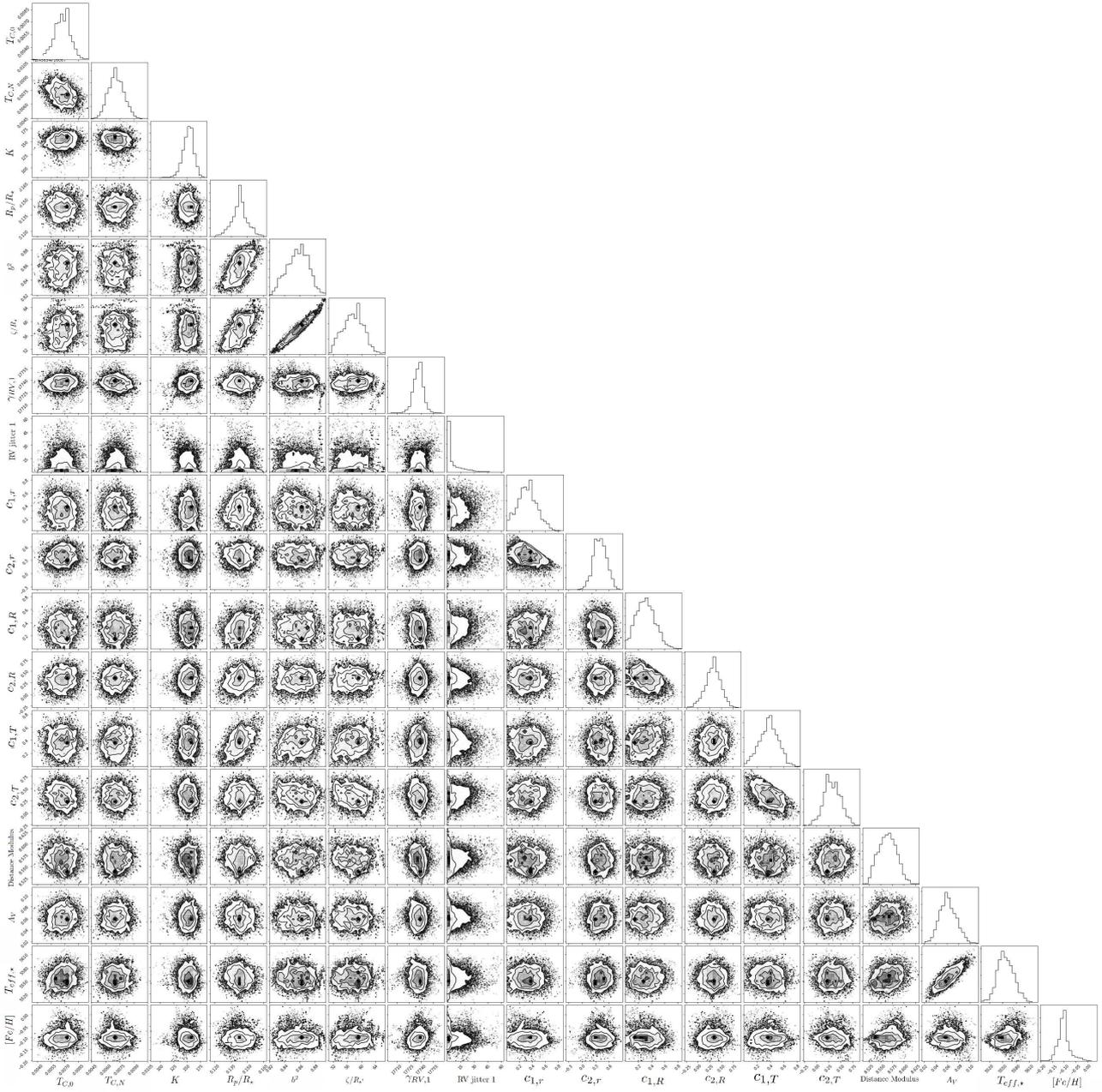

**Figure A1. Corner Plot for HATS-34** This figure shows a corner plot for the parameters varied for the isochrone-based fit of HATS-34.

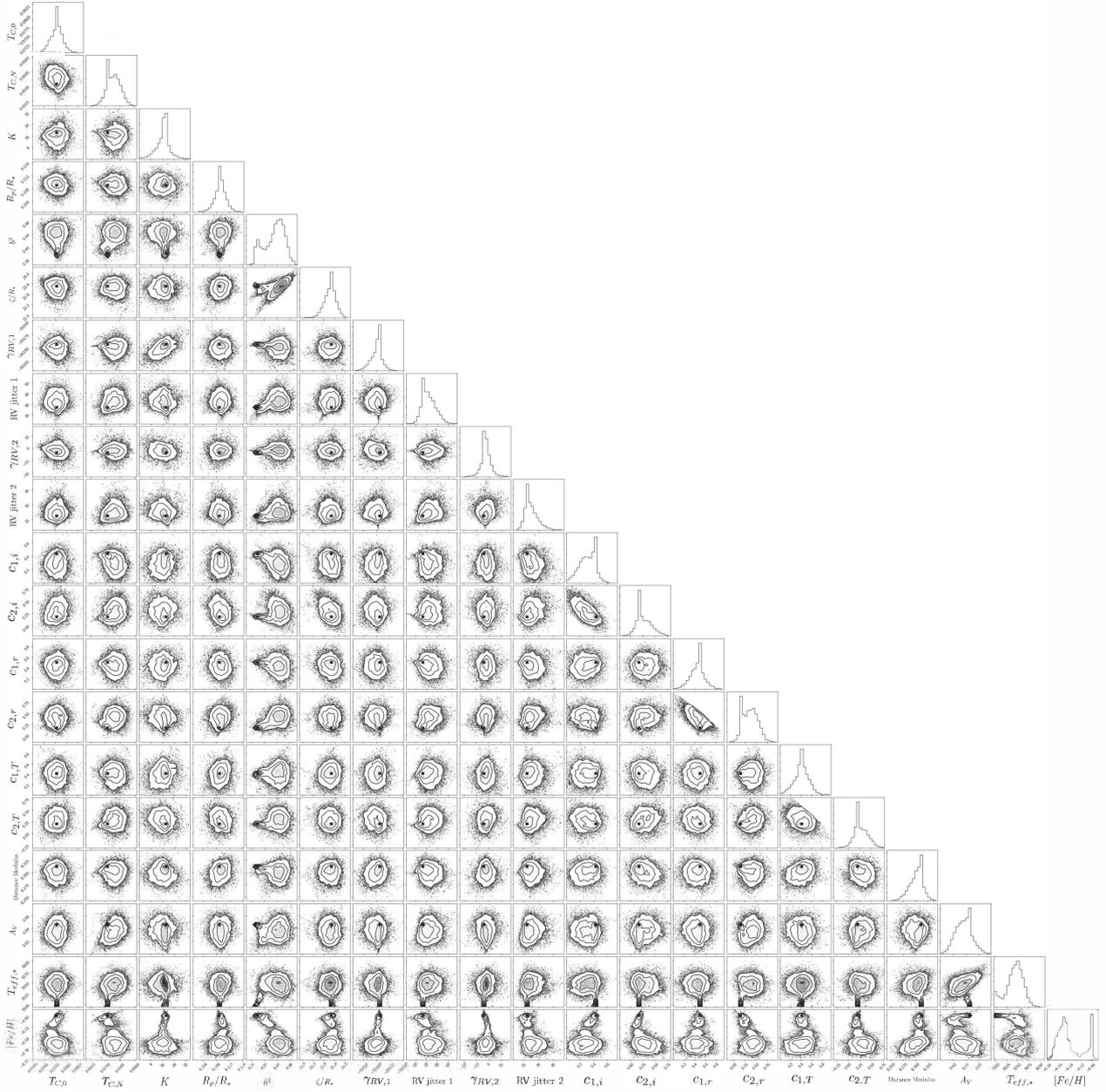

**Figure A2. Corner Plot for HATS-46** This figure shows a corner plot for the parameters varied for the isochrone-based fit of HATS-46.